\documentclass[%
reprint,
 amsmath,amssymb,
 aps,
 prl,
]{revtex4-2}

\usepackage{graphicx,import}
\usepackage{dcolumn}
\usepackage{mathtools}
\newcommand*\dif{\mathop{}\!\mathrm{d}} 
\usepackage{siunitx}
\usepackage[thinc]{esdiff}
\usepackage{bm}
\usepackage{svg}
\usepackage{lipsum}
\usepackage{textcomp}
\usepackage{hyperref}

\begin{document}

\title{Liquid Surfaces with Chaotic Capillary Waves Exhibit an Effective Surface Tension}

\author{Steffen Bisswanger}
\author{Henning Bonart}%
\author{Pyi Thein Khaing}
 \author{Steffen Hardt}%
 \email{hardt@nmf.tu-darmstadt.de}
\affiliation{%
Technische Universität Darmstadt, Institute for Nano- and Microfluidics, Peter-Grünberg-Straße 10, D-64287 Darmstadt, Germany
}%


\date{\today}

\begin{abstract}
The influence of chaotic capillary waves on the time-averaged shape of a liquid volume is studied experimentally and theoretically. In that context, a liquid film containing a stable hole is subjected to Faraday waves. The waves induce a shrinkage of the hole compared to the static film, which can be described using the Young-Laplace equation by incorporating an effective capillary length. In the regime of chaotic Faraday waves, the presented theory explains the hole shrinkage quantitatively, linking the effective capillary length to the wave energy. The effect of chaotic Faraday waves can be interpreted as a dynamic surface force that acts against  surface tension.
\end{abstract}

\maketitle

\noindent Out-of-equilibrium liquid surfaces exhibit very rich and intriguing dynamics. A common way to excite dynamic modes on a liquid surface is to vibrate a liquid volume. The mean (time-averaged) free surface of a vibrated liquid volume can take shapes that are significantly different from that in equilibrium~\cite{Beyer.2001, Gavrilyuk.2004}. Further, by vibrating a liquid, its stability behavior can be influenced and unstable modes can be eliminated in specific parameter regimes. For example, the Rayleigh-Taylor instability of superposed liquids can be suppressed by vibrating the system normal to the interface~\cite{Wolf.1970}, or a liquid bridge arranged vertically between two disks can be stabilized by vibrating the upper disk~\cite{Haynes.2018}. Vibrating an entire liquid volume is not the only way to excite dynamic modes on a liquid surface. Surprisingly, when exposing a sessile drop to ultrasound, capillary waves are observed on the surface of the drop that have a much lower frequency than the excitation source \cite{Qi.2008, Tan.2010}. A few attempts have been made to explain the nonlinearities that cause the corresponding energy transfer from high to low frequency modes~\cite{Blamey.2013, Zhang.2023}. 

A classical configuration used to create capillary waves on a liquid is to vibrate a liquid pool at a comparatively low frequency of a few 100 Hz or smaller. Above a critical acceleration amplitude, so-called Faraday waves are observed ~\cite{Faraday.1830, Benjamin.1954}, a phenomenon that has been studied intensely during the past decades. The classical configuration, however, gives little information about how such surface waves can affect the time-averaged shape of a liquid volume. Corresponding information can be obtained when a drop is deposited on a liquid pool that is vibrated, such that Faraday waves can be excited on the drop surface, while the surface of the pool remains below the excitation threshold~\cite{Pucci.2011, Pucci.2015}. Two regimes are observed: A quasi-steady regime in which the drop takes a new (time-averaged) equilibrium shape and a second unsteady regime in which the drop shape exhibits large variations in time. One fundamental challenge related to capillary waves is the question whether the complex dynamics of such systems can be captured by simplified theoretical descriptions. In the simplest case, important aspects of the dynamics would be captured by mapping the system to a stationary equilibrium system with effective parameters. An important step in that direction was made by Welch et al. who studied chaotic Faraday waves on a liquid pool~\cite{Welch.2016, Welch.2014}. They showed that a probe immersed in the liquid behaves as if the liquid has an effective temperature and viscosity. These quantities are tunable via the frequency and amplitude of the shaker.

\begin{figure*}
    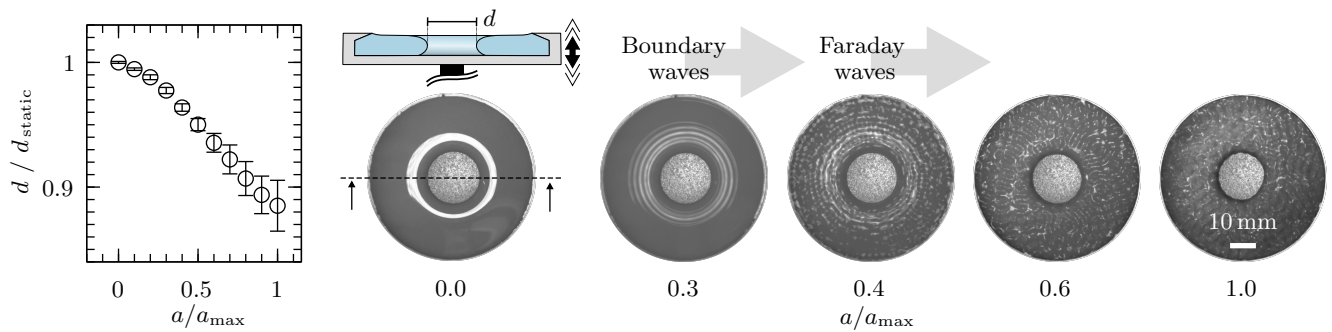
    \caption{\label{fig:d_over_a}Stable holes in vibrated liquid films shrinking with increasing excitation amplitude. The plot on the left shows the normalized time-averaged hole diameter as a function of the dimensionless acceleration amplitude and corresponds to the images on the right. The error bars show the standard deviation for subsequent measurements. $f_{\mathrm{ex}}=\qty{170}{\hertz}$, $d_{\mathrm{static}}$=\qty{24.6}{\milli\metre} and $a_{\mathrm{max}}=\qty{68.36}{\metre\per\square\second}$.}
\end{figure*}

\begin{figure}[h!]
    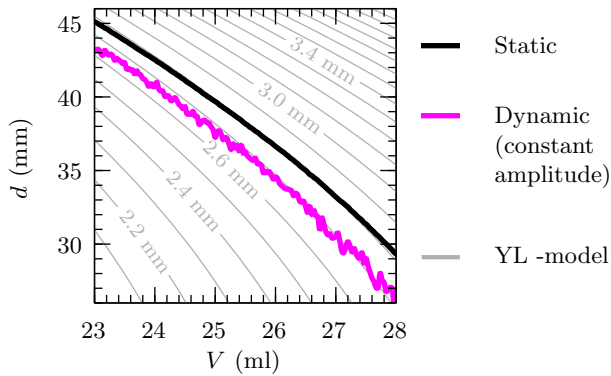
	\caption{Evolution of the hole diameter with liquid volume in the film. The black and magenta curves show experimental data obtained from a static film and a vibrated one, respectively. Thin gray lines represent numerical solutions of the Young-Laplace equation for different capillary lengths. $f_{\mathrm{ex}}=\qty{116}{\hertz}$, $A_{\mathrm{ex}}\omega_{\mathrm{ex}}=\qty{0.056}{\metre\per\second}$}\label{fig:d_over_V}
\end{figure}

In the present work, we consider a similar situation as in~\cite{Welch.2016}, but focus on surface properties of the liquid instead of bulk properties. Specifically, we will show that a liquid volume on which chaotic Faraday waves are excited behaves as if it possesses an effective surface tension that can be tuned via the frequency and amplitude of the shaker. In that context, the dynamic system can be mapped to a static equilibrium system that is described by the Young-Laplace equation. 
Consider a bounded film of water on a horizontal surface. In this configuration we can create a defect or hole in the liquid film that is stable~\cite{Sharma.1990}.
Above a critical radius we can vary the size of the hole by changing the liquid volume inside the container~\cite{Lv.2018}. Using a superhydrophobic substrate~\cite{Gupta.2016} we obtain a rather thick film ($h\approx\qty{5}{\milli\metre}$) that is very sensitive to disturbances by external forces. This makes it an ideal candidate to study the effect of capillary waves on the time-averaged shape of the free-surface.
When subjecting this system to vertical vibrations, as shown in Fig.~\ref{fig:d_over_a}, we see concentric patterns of boundary waves for low excitation amplitudes. When surpassing a critical amplitude we see Faraday waves that become more chaotic with increasing amplitude. The time-averaged diameter of the hole shrinks continuously with increasing amplitude.
In a second set of experiments we keep the vibration amplitude constant and instead slowly add liquid to the film. The resulting hole diameter is shown in Fig.~\ref{fig:d_over_V}. Remarkably, the dynamic system evolves very similarly as the static one. The hole shrinks with increasing volume. Using the Young-Laplace equation we can describe how the hole in the static film should evolve in this $d$-$V$-space. Comparing the Young-Laplace model with the experimental results it seems as if the dynamic system behaves as if it has an effective capillary length. In combination with the results shown in  Fig.~\ref{fig:d_over_a} it is natural to assume that the effective capillary length also depends systematically on the wave dynamics. This raises two important questions: How exactly is this effective capillary length related to the wave dynamics, and is there a physically meaningful interpretation of it?
\\\noindent To answer these questions, we start with the well established model for the static axisymmetric configuration, similar as in~\cite{Lv.2018}. 
The pressure jump across the gas-liquid interface is described by the Young-Laplace equation and balances the hydrostatic pressure inside the liquid film
\begin{align}
    -\sigma\left(\diff{\phi}{s} + \frac{\sin{\phi}}{r}\right) &= -\rho g z + \Delta p_0,
    \label{eq:YL}
\end{align}
where surface tension, mass density and gravity are denoted by $\sigma$, $\rho$ and $g$, respectively. The radial and vertical coordinates are $r$ and $z$, where $z(r)$ describes the liquid surface. The arc-length coordinate along the liquid surface is denoted by $s$, the local inclination angle of the surface by $\phi$. Additionally, we have $\mathrm{d}r/\mathrm{d}s=\cos{\phi}$ and $\mathrm{d}z/\mathrm{d}s=\sin{\phi}$. The Laplace pressure at the solid surface ($z=0$) is denoted by $\Delta p_0$. The boundary conditions are $\phi|_{s=0}=\theta$, $r|_{s=0}=r_0$, $z|_{s=0}=0$ and $z|_{r=R}=z_{R}$, where $\theta$, $r_0$, and $R$ are the contact angle, hole radius and container radius, respectively. The latter boundary condition corresponds to pinning the liquid film to a sharp edge at the perimeter (c.f. Fig.~\ref{fig:d_over_a}). The problem can be non-dimensionalized to depend only on the contact angle and the capillary length $l_c = \sqrt{\sigma/\rho g}$.

\begin{figure}[h!]
	\includegraphics{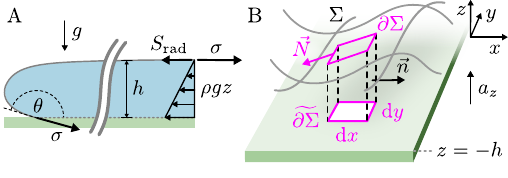}
	\caption{Liquid puddle model introducing the effective capillary length (A), and control volume in the deformed liquid film considered to compute $S_{\mathrm{rad}}$ (B). Only forces that act in the horizontal direction are shown.}\label{fig:lc_eff}
\end{figure}

\noindent{}
To describe the dynamics of the film with capillary waves, we introduce an effective capillary length $l_c^\mathrm{eff}$ and aim at modeling its time-averaged configuration using a static surrogate system described by equation \ref{eq:YL}. For simplicity, we neglect the second curvature term $\sin(\phi)/r$ in equation \ref{eq:YL}, which is justified for holes that are large compared to the film height. As a consequence, we can resort to a 2D liquid puddle, as shown in Fig.~\ref{fig:lc_eff}.
The liquid puddle height $h$ depends on the hydrostatic pressure inside the film, surface tension and the contact angle. Balancing the forces in horizontal direction gives
\begin{align}
    h = \sqrt{2(1-\cos{\theta})\cdot l_c^{\:2}}\label{eq:liquid_puddle_height}
\end{align}
for the static liquid puddle~\cite{deGennes.2004}. In order to apply this model to the dynamic system with capillary waves, we need to compute the effective capillary length. We introduce a horizontal force $S_{\mathrm{rad}}$ (c.f. Fig.~\ref{fig:lc_eff} A) that acts in addition to surface tension, far away from the meniscus, and represents the dynamic contribution of the waves. Considering the horizontal force balance at the liquid surface, we obtain a modified expression for the capillary length:
\begin{align}
    l_c^{\mathrm{eff}} = \sqrt{l_c^2 - \frac{S_{\mathrm{rad}}}{\rho g(1-\cos{\theta})}}.\label{eq:lc_eff}
\end{align}
\\\noindent Next, we study how $S_{\mathrm{rad}}$ depends on the waves in the liquid film.
Analyses on the excess momentum flux in capillay wave systems have been performed before, e.g. by Longuet-Higgins and Stewart \cite{LonguetHiggins.1964}. In analogy to the pressure that electromagnetic waves exert onto surfaces, this additional momentum flux is often called radiation pressure or radiation stress in the anisotropic case. This radiation pressure can be related to the area specific energy contained in the waves. This is explained in detail in the Supplementary Material~\cite{SM}. The most important steps in the calculation of $S_{\mathrm{rad}}$ are outlined in the following. 

\begin{figure}
    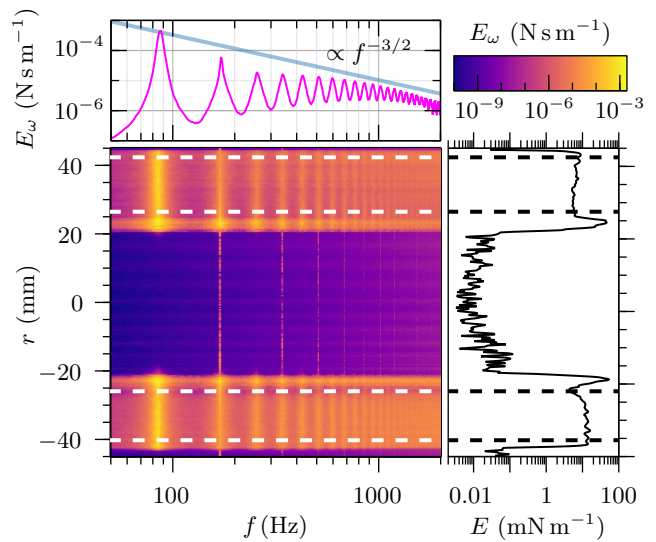
	\caption{Experimental results for the area-specific wave energy. The colormap shows the spatially resolved energy spectrum of the vibrated film with a hole at the center. The vertical axis indicates different measurement positions along the radial direction. At the top, the spatially averaged wave spectrum is shown. The blue line indicates the scaling prediction from capillary wave turbulence theory. The total spatially resolved wave energy $E$ is shown on the right. $f_{\mathrm{ex}}=\qty{170}{\hertz}$, $A_{\mathrm{ex}}\omega_{\mathrm{ex}}=\qty{57}{\milli\metre\per\second}$}\label{fig:wave_energy}
\end{figure}

We consider the system shown in Fig.~\ref{eq:lc_eff} B that is based on an infinitely extended film. We parameterize the surface $\Sigma$ by a height function $z = \eta(x,y,t)$ that describes the displacement relative to the static film for which we have $z=0$. The liquid film is subject to a time-harmonic vertical acceleration 
\begin{align}
    a_{z} = - g + A_{\mathrm{ex}}\omega_{\mathrm{ex}}^{2}\cos{\left(\omega_{\mathrm{ex}}t\right)}.
\end{align}
We assume that the induced waves are chaotic such that
\begin{align}
   \left\langle\eta(x,y,t)\right\rangle_t = 0,\label{eq:surf_average}
\end{align}
where $\langle\cdot\rangle_{t}$ denotes local time averaging over a period much larger than a wave period. 
Further, we assume that for the dynamic system $\partial/\partial{}t\left\langle\cdot\right\rangle_t=0$ holds for all field quantities. Therefore, for the dynamic system the time-averaged momentum flux into the volume shown in Fig.~\ref{fig:lc_eff} B is zero. Assuming the flow to be inviscid, the $x$-momentum balance over the control volume reads
\begin{align}
0 = \left\langle\:\int\displaylimits_{\partial V}\hspace{-0.5em}\left(\rho uvn_y +\rho u^2n_x+ pn_x\right) \dif A -\int\displaylimits_{\partial\Sigma}\hspace{-0.4em}\sigma N_x \dif s\right\rangle_t,\label{eq:flux_into_control_volume}
\end{align}
where $u,v$ denote the $x,y$-components of the velocity field, respectively, $p$ pressure, $n_x, n_y$ the components of the outward unit normal vector on the control volume, and $N_x$ is the $x$-component of the tangent unit vector at the liquid surface normal to $\partial\Sigma$.
Assuming the waves and the velocity field to be statistically isotropic in the horizontal plane the shear-stress contribution from the first term in the integral vanishes. It also implies that we do not loose generality by examining only the $x$-direction. Using the small angle approximation for the liquid surface we obtain $N_x=\cos\left(\partial\eta/\partial x\right)$ and $\mathrm{d}y=\cos\left(\partial\eta/\partial y\right)\mathrm{d}s$. Splitting the integration over $\partial{}V$ into one over the time-independent line $\widetilde{\partial\Sigma}$ (c.f. Fig.~\ref{fig:lc_eff}B) and the film height, we pull the time averaging in equation \ref{eq:flux_into_control_volume} under the time-independent line integral, giving
\begin{align}
    0 = \int\displaylimits_{\widetilde{\partial\Sigma}} \left\langle\Phi_{\mathrm{dyn}}\right\rangle_t n_x \dif{}\tilde{s}
\end{align}
where
\begin{align}
\left\langle\Phi_{\mathrm{dyn}}\right\rangle_t = \left\langle\:\int_{-h}^{\eta}\left(\rho u^2+ p\right) \dif z -\sigma\frac{\cos{\left(\frac{\partial\eta}{\partial x}\right)}}{\cos{\left(\frac{\partial\eta}{\partial y}\right)}}\right\rangle_t\label{eq:momentum_flux_density}
\end{align}
is the height integrated, time averaged horizontal momentum flux per unit area.
We define the radiation pressure $S_{\mathrm{rad}}$ as the difference in horizontal momentum flux between the time-averaged dynamic system and the static system
\begin{equation}
\begin{split}
    S_{\mathrm{rad}} \coloneqq& \left\langle\Phi_{\mathrm{dyn}}\right\rangle_t
    -\left(\int_{-h}^{0}-\rho gz\dif z-\sigma\right).
    \label{eq:Srad_def}
\end{split}
\end{equation}
Exploiting the separation of timescales between the oscillation period and the chaotic motion of the wave patterns, we can assume standing waves. Then we have $\left\langle\cos(\partial\eta/\partial x)/\cos(\partial\eta/\partial y)\right\rangle_t = 1$. To further simplify equation \ref{eq:Srad_def}, we split the pressure integral in two at $z=\eta_{\mathrm{min}}=\min{(\eta(x,y,t))}$. Then equation \ref{eq:Srad_def} becomes $S_{\mathrm{rad}} = S_{\mathrm{rad}}^{(1)} + S_{\mathrm{rad}}^{(2)}$ with
\begin{equation}
\begin{split}
    S_{\mathrm{rad}}^{(1)} \coloneqq& \left\langle\int_{-h}^{\eta} \rho u^2\dif z\right\rangle_t+\int_{-h}^{\eta_\mathrm{min}}\left(\left\langle p\right\rangle_t+\rho gz\right)\dif z\label{eq:Srad_1_def}
\end{split}
\end{equation}
and
\begin{equation}
\begin{split}
    S_{\mathrm{rad}}^{(2)} \coloneqq& \left\langle\int_{\eta_\mathrm{min}}^{\eta}p\dif z\right\rangle_t+\int_{\eta_\mathrm{min}}^{0}\rho gz \dif z.\label{eq:Srad_2_def}
\end{split}
\end{equation}
We employ a time-averaged vertical momentum balance to reformulate $\left\langle p\right\rangle_t$. Further, we assume that $\eta$ does not have a significant frequency component at $\omega_{\mathrm{ex}}$, that the flow is curl-free and that velocity gradients close to the interface are small, such that the pressure inside the first integral of equation \ref{eq:Srad_2_def} is $p=\rho g(\eta -z) -\sigma\kappa$, where we define the sign of the curvature $\kappa$ so that $\kappa<0$ when the liquid film is locally convex. Equations \ref{eq:Srad_1_def} and \ref{eq:Srad_2_def} can then be rewritten as
\begin{equation}
\begin{split}
    S_{\mathrm{rad}}^{(1)} =& \left\langle\int_{-h}^{\eta}\rho u^2\dif z\right\rangle_t - \int_{-h}^{\eta_{\mathrm{min}}}\rho\left\langle w^2\right\rangle_t\dif z \label{eq:Srad_1_reform}
\end{split}
\end{equation}
and
\begin{equation}
\begin{split}
    S_{\mathrm{rad}}^{(2)} =& \frac{1}{2}\rho g\left\langle\eta^2\right\rangle_t-\sigma\left\langle\kappa\eta\right\rangle_t,\label{eq:Srad_2_reform}
\end{split}
\end{equation}
respectively. For standing waves, the correlation term in equation \ref{eq:Srad_2_reform} containing the curvature and film height can be related to the wave number as $\left\langle\kappa\eta\right\rangle_t=k^2\left\langle\eta\right\rangle_t/2$. We can now see that $S_{\mathrm{rad}}^{(2)}$ is equal to the potential energy of the waves. Assuming equipartition between kinetic and potential energy ~\cite{Galtier.2021}, we have $S_{\mathrm{rad}}^{(1)}\ll S_{\mathrm{rad}}^{(2)}$. Introducing the Bond number $Bo=\rho g/\sigma k^2$ and neglecting $S_{\mathrm{rad}}^{(1)}$, we finally arrive at 
\begin{align}
    S_{\mathrm{rad}} = \frac{1}{2}\rho g\left\langle\eta^2\right\rangle_t\left(1+Bo^{-1}\right)=\frac{E}{2},\label{eq:S_rad_on_E}
\end{align}
where $E$ is the total wave energy per unit area.
For $Bo^{-1}\gg 1$, the main contribution to the radiation pressure stems from the term $\sigma\left\langle\kappa\eta\right\rangle_t$ in equation \ref{eq:Srad_2_reform}, i.e. the fact that that pressure fluctuations close to the interface caused by the Laplace pressure are locally correlated with the height of the liquid film. We can trace the origin of the radiation pressure further back to equation \ref{eq:Srad_2_def}, showing that for $|\eta_{\mathrm{min}}|/h\ll 1$, this contribution is restricted to regions close to the surface. Correspondigly, we interpret the radiation pressure as a dynamic surface force that acts against surface tension.
The model presented above implies the possibility to predict the hole shrinkage in the vibrated liquid film solely based on the wave energy. 
For the experimental validation we choose a frequency range of $\qty{100}{\hertz}\leq f_{\mathrm{ex}} \leq \qty{200}{\hertz}$, with velocity amplitudes of $0 \leq A_{\mathrm{ex}}\omega_{\mathrm{ex}} < \qty{65}{\milli\metre\per\second}$. This effectively avoids radial oscillation modes and splashing waves, and covers the interval of the expected applicability of our theory. Using the inviscid dispersion relation $\omega^2 = \tanh{(kh)}\left(gk + (\sigma k^3)/\rho\right),\label{eq:disp_rel}$ we obtain $Bo^{-1} > 8$ for all subharmonic waves. The damping parameter for subharmonic waves is $\gamma = 2\nu k_0^2/\omega_0< 0.01$. The normalized supercritical amplitude is  $\varepsilon\coloneqq(A_{\mathrm{ex}}\omega_{\mathrm{ex}}^{2}-a_c)/a_c< 4$, where $a_c$ is the critical amplitude.
We determine $S_{\mathrm{rad}}$ in two independent ways: First, we follow our hypothesis that we can infer the radiation pressure based on the hole shrinkage. We image the hole in the liquid film from above and determine the diameter of the hole.
We record the hole diameter for different amplitudes, frequencies and liquid volumes. 
Then we estimate the effective capillary lengths of the vibrated systems using a Bayesian hierarchical model involving the Young-Laplace equation via variational inference.
Using equation~\ref{eq:lc_eff}, this gives an estimate for $S_{\mathrm{rad}}$ including credibility intervals.
Second, we measure the potential energy of the waves and relate it to $S_{\mathrm{rad}}$ via equation~\ref{eq:S_rad_on_E}. 
To obtain the potential energy we employ a spectral method, which is necessary since a substantial amount of wave energy is found at frequencies above the subharmonic mode. Using a laser sheet triangulation sensor we record a time series of the film height $\eta(x_0,y_0,t)$ and determine its power spectral density $\left\langle\eta_{\omega}^2\right\rangle_t$. 
Similarly as in~\cite{Berhanu.2013, Deike.2014, Berhanu.2018}, we obtain the wave energy spectrum $E_{\omega}$ by 
\begin{align}
    E_{\omega} = \rho g\left\langle\eta_{\omega}^2\right\rangle_t\left(1+Bo^{-1}\right).
\end{align}
The resulting energy spectrum for one parameter combination is shown in figure \ref{fig:wave_energy}. Via the dispersion relation, the Bond number is a function of $\omega$. The wave energy is given as
\begin{align}
    E = \frac{1}{2\pi}\int_0^{\omega_{\mathrm{cut}}}E_{\omega}\dif\omega.
\end{align}
Weak capillary wave turbulence theory predicts a scaling of $E_{\omega}\propto f^{-3/2}$ (c.f. Fig.~\ref{fig:wave_energy})~\cite{Zakharov.1967}. Therefore, the disregarded energy beyond $\omega_{\mathrm{cut}} = \qty{1440}{\hertz}$ introduces a relatively small error. For details on the wave energy measurement, the experimental setup and estimation on $S_{\mathrm{rad}}$ the reader is referred to the Supplementary Material~\cite{SM}  (see also references~\cite{Gupta.2016, Galtier.2021, Hatatani.2022, Huang.2021, Przadka.2012, Deike.2014, Zakharov.1967, LonguetHiggins.1964} therein).

\begin{figure}
	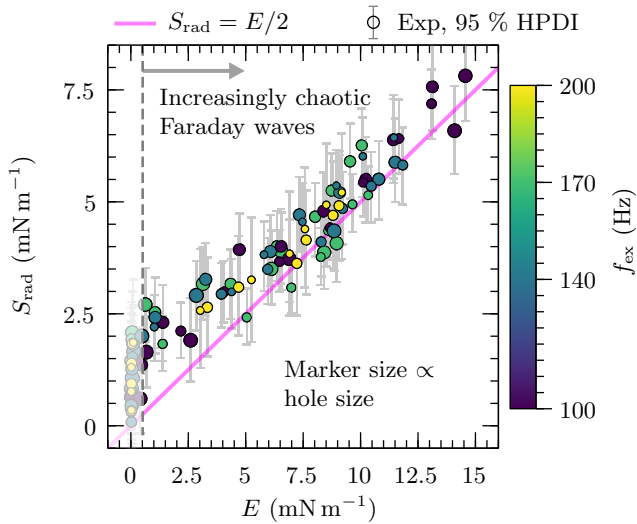
	\caption{Comparison between measurement of $S_{\mathrm{rad}}$ via wave energy and estimation of $S_{\mathrm{rad}}$ according to our hypothesis on hole shrinkage. The symbols represent the experimental data, with the marker size and color indicating the hole diameter and excitation frequency, respectively. The error bars correspond to the $\qty{95}{\percent}$ highest posterior density interval (HDPI), see Supplementary Material for details~\cite{SM}. The magenta line indicates the theoretical prediction. The area left of the dashed line corresponds to boundary waves for which the theory is not applicable. }\label{fig:srad_over_e_log_lin_sc}
\end{figure}

Fig.~\ref{fig:srad_over_e_log_lin_sc} shows how $S_{\mathrm{rad}}$ (obtained from hole shrinkage) correlates with $E$.  
The prediction according to equation \ref{eq:S_rad_on_E} is indicated by the magenta line. As expected, we do not find agreement for low wave energies, where we only have harmonic boundary waves, so that the theory is not applicable. However, when surpassing energies at which Faraday waves start to emerge, we see an increasing agreement between theory and experiments. At large wave energies of  $E>\qty{5}{\milli\newton\per\metre}$ we see excellent quantitative agreement. In accordance with the theory, the experimental data do not show any systematic dependency on the hole size (c.f. marker size in Fig.~\ref{fig:srad_over_e_log_lin_sc}) or frequency (c.f. color grading in Fig.~\ref{fig:srad_over_e_log_lin_sc}). The experimental range extends up to wave energies of $E\approx\qty{15}{\milli\newton\per\metre}$ i.e. a radiation pressure of $S_{\mathrm{rad}}\approx\qty{7.5}{\milli\newton\per\metre}$. This corresponds to an effective reduction of the surface tension of more than $\qty{10}{\percent}$. Using this upper bound for the energy and $Bo^{-1}>8$, we obtain $\sqrt{2\left\langle\eta^{2}\right\rangle_t}/h<0.11$. Thus, we interpret $S_{\mathrm{rad}}$ as a dynamic surface force. 
In conclusion, we presented a theory that connects the dynamics of a vibrated liquid film to its static behavior. The surface waves on the film cause a radiation pressure than can be understood as an effective contribution to surface tension. We conducted experiments that are in quantitative agreement with the theory within its expected range of validity. As surface tension can play a pivotal role in the stability of liquid volumes, our findings could explain the stabilizing or destabilizing effects of capillary waves. More generally, the concept of an effective surface tension of vibrated liquids could open the door to far-reaching analogies, such as Marangoni stresses that arise due to variations of the wave energy.  
\section{Acknowledgements}
We wish to acknowledge the help by Matthias Weigold, TU Darmstadt, who provided the Laser triangulation sensor for the wave energy measurement.
Funding by the Deutsche Forschungsgemeinschaft (DFG, German Research Foundation), Project IDs 459970814, 459970841 and 455566770 is greatly acknowledged. S.B. likes to thank Salar Jabbary Farrokhi for helpful discussions.


\bibliography{literature}

\end{document}



\title{Liquid Surfaces with Chaotic Capillary Waves Exhibit an Effective Surface Tension\\\vspace{0.5cm}Supplementary Material}

\author{Steffen Bisswanger}
\author{Henning Bonart}%
\author{Pyi Thein Khaing}
 \author{Steffen Hardt}%
 \email{hardt@nmf.tu-darmstadt.de}
\affiliation{%
Technische Universität Darmstadt, Institute for Nano- and Microfluidics, Peter-Grünberg-Straße 10, D-64287 Darmstadt, Germany
}%


\date{\today}


\maketitle

\tableofcontents
\newpage\section{Experiments}
\subsection{Setup}
\subsubsection{Properties of the Liquid}
\noindent For all experiments a $\mathrm{TiO}_{2}$-water dispersion is used to enable the laser triangulation sensor to detect the height profile of the gas-liquid interface. The particles used are of crystalline type anatase in order to minimize the effects on the capillary waves compared to pure water as suggested by Przadka et al. ~\cite{Przadka.2012}. The particle size is $\qty{200}{\nano\metre}$ to $\qty{400}{\nano\metre}$. The batch purity is $\qty{99.5}{\percent}$ (\textit{Carl Roth} Titanium(IV) oxide, Item No. 9985, Certificate of Analysis for batch no. 083334403). The dispersion is prepared by adding $\qty{0.1}{\gram}$ of particles to $\qty{100}{\milli\litre}$ of \textit{milli-Q} water. Before usage, the mixture is stirred for 10 minutes and then agitated in a ultrasonic water bath for 10 minutes to properly disperse the particles. The viscosity of the dispersion and pure water as a reference is measured using a cone rheometer (\textit{Brookfield} model \textit{DV-III Ultra}). The determined values are shown in table \ref{tab:liquid_prop}.
The surface tension is determined by means of pendant drop tensiometry using a \textit{Krüss} model \textit{DSA 100} for imaging and the open source software \textit{OpenDrop} for evaluation~\cite{Huang.2021}. The determined values are shown in table~\ref{tab:liquid_prop}. 
\noindent With a volumetric fraction of $\mathrm{TiO}_2$ of
\begin{align}
    \phi = \frac{\qty{1}{\gram}/(\qty{4.26}{\gram\per\milli\litre})}{\qty{1000}{\milli\litre}+\qty{1}{\gram}/(\qty{4.26}{\gram\per\milli\litre})} = 2.3\cdot 10^{-4},
\end{align}
the Einstein relation predicts an increase in viscosity of $5/2\cdot 2.3\cdot 10^{-4} = \qty{0.06}{\percent}$ for the dispersion compared to pure water. This is consistent with the measurements listed in table~\ref{tab:liquid_prop}, considering the specified standard deviation (SD).
\subsubsection{Properties of the Substrate}
\begin{figure}[!h]
    \includegraphics[width=\textwidth]{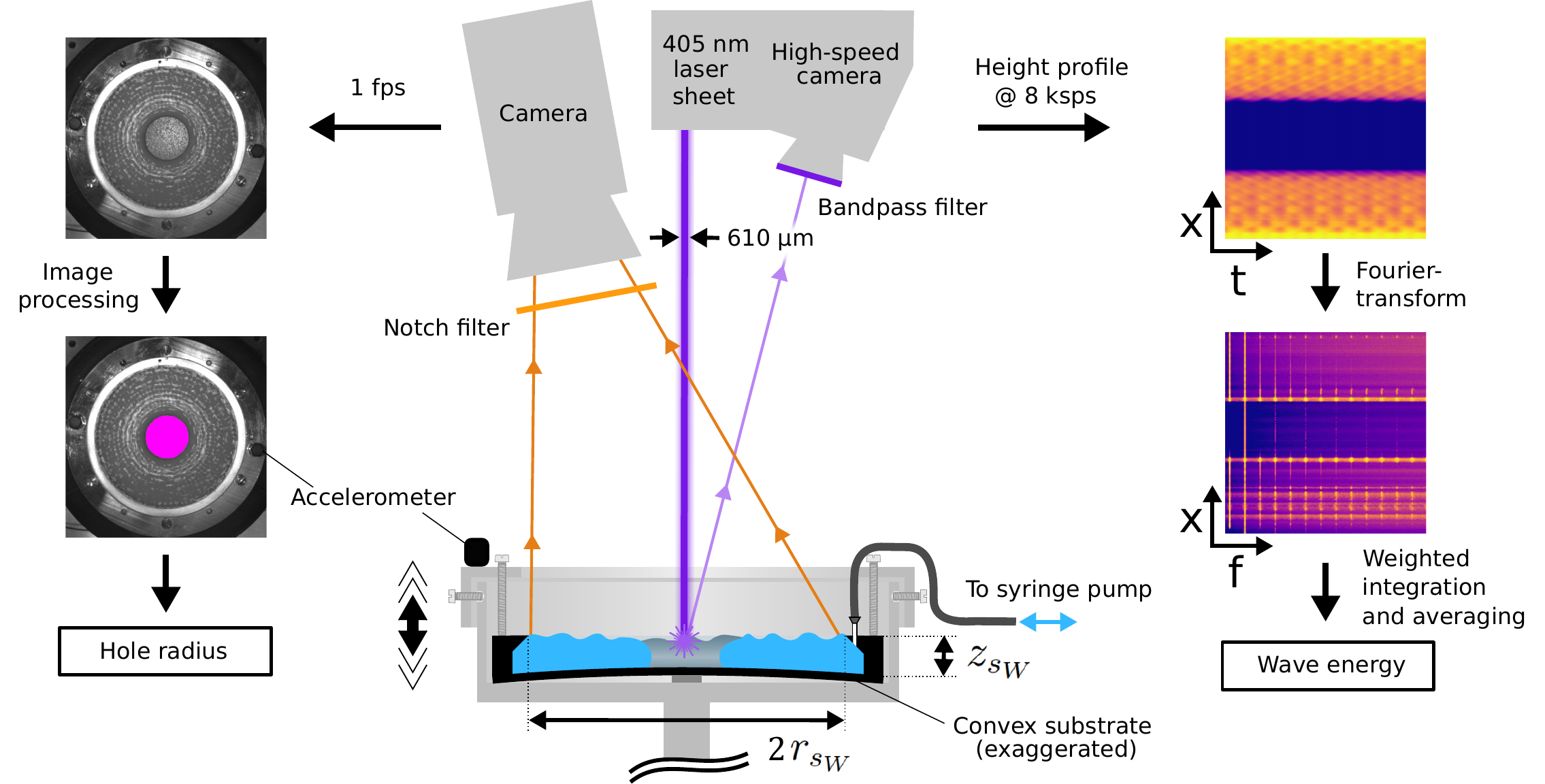}
    \caption{\justifying Setup for the simultaneous measurement of hole radius and wave energy. The basin is mounted on a vibration exciter (\textit{Bruel \& Kjaer} model \textit{4809}). It is driven by a control unit (\textit{Bruel \& Kjaer} model \textit{LDS COMET USB}) via a power amplifier (\textit{Bruel \& Kjaer} model \textit{2706}). The acceleration is closed-loop controlled using a piezoelectric sensor (\textit{Bruel \& Kjaer} model \textit{4519-002}). The liquid volume inside the basin is controlled by a syringe pump (\textit{KD Scientific} model \textit{200}) using a 50 ml syringe. Similarly as in \cite{Hatatani.2022}, the deformation of the water surface is recorded using a laser profilometer (\textit{Keyence} model \textit{LJ-7300}). It projects a laser sheet ($\qty{405}{\nano\metre}$ wavelength; $\qty{610}{\micro\metre}$ thickness) such that it is normal to the basin. Imaging it from one side gives the height profile of the liquid surface by means of triangulation. The pixel spacing on the image sensor results in a horizontal spacing of \qty{300}{\micro\metre} for the height profile data. The sampling rate is set to 8000 samples per second (8 ksps). The imaging for the hole size measurement is done from above at almost 90°, resulting in no noticeable perspective distortion. A notch filter in front of the camera is used to suppress the light of the $\qty{405}{\nano\metre}$ laser on the images that are recorded at 1 fps.}
    \label{fig:visual_setup}
\end{figure}

\noindent For the fabrication of the substrate, a $\qty{380}{\micro\metre}$ thick 4-inch Si wafer is treated with the commercially available superhydrophobic coating \textit{NeverWet Rustoleum} (frosted clear)~\cite{Gupta.2016}. It is applied manually in several base coats forming a polymer film and subsequent top coats covering it with hydrophobic nano particles, making the substrate superhydrophobic. The measured values for the advancing and receding contact angle on one sample substrate are $\theta_{\mathrm{a}}=163^{\circ}\pm 5^{\circ}$ and $\theta_{\mathrm{r}}=152^{\circ}\pm 6^{\circ}$, respectively.
The coating has a thickness of $(53\pm 7)\qty{}{\micro\metre}$, measured using a micrometer screw gauge across different locations on different substrates. To allow for centering the hole in the liquid film, the substrate has to have a slightly raised center area compared to the surrounding area that is covered by the liquid film. To achieve this, a circular patch ($D=\qty{8}{\milli\metre}$, $h=\qty{40}{\micro\metre}$) of adhesive tape is placed at the center below the substrate. When pressing the substrate down onto the basin (c.f. Fig.~\ref{fig:visual_setup}), it is forced into a slightly convex shape. A height profile measurement of the clamped substrate with the resulting shape is shown in Fig.~\ref{fig:substrate_concavity}. Compared to the height of the liquid film of about 5 mm, the substrate is comparatively flat. In order to avoid trapping air during the filling process of the basin at the edge where the substrate meets the bounding wall, the substrate is left uncoated, i.e. hydrophilic, close to the bounding wall. A schematic overview of the entire setup is shown in Fig. \ref{fig:visual_setup}.
\begin{minipage}{\textwidth}
  \begin{minipage}[b]{0.4\textwidth}
    \centering
    \includegraphics[width = \textwidth]{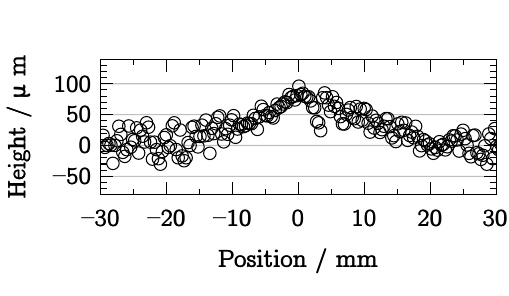}
    \captionof{figure}{Height profile of the clamped substrate. The profile is measured with the laser triangulation sensor. The data do not  extend to the outer edge, where the missing coating inhibits a measurement by the laser sensor of the bare silicon wafer surface.}\label{fig:substrate_concavity}
  \end{minipage}
  \hfill
  \begin{minipage}[b]{0.55\textwidth}
    \centering
\setlength{\tabcolsep}{10pt} 
\renewcommand{\arraystretch}{1.05} 
\footnotesize
\begin{tabular}{lcccc}
\hline\hline\\
                     & \multicolumn{2}{c}{$\nu / $cSt} & \multicolumn{2}{c}{$\sigma / \qty{}{\milli\newton\per\metre}$} \\
                     & mean         & SD       & mean                        & SD                       \\ \hline\\
pure water           & 1.061        & $\pm 0.013$            & 74.8                        & $\pm 0.5$                              \\
$\qty{1}{\gram\per\litre}$ $\mathrm{TiO}_2$ & 1.074        & $\pm 0.031$            & 74.3                        & $\pm 0.4$                              \\ \\\hline\hline
\end{tabular}\label{tab:liquid_prop}\vspace{10pt}
      \captionof{table}{Comparison of measured properties of pure water and the $\qty{1}{\gram\per\litre}$ $\mathrm{TiO}_2$ dispersion.}\label{tab:liquid_prop}
      \vspace{55pt}
    \end{minipage}
  \end{minipage}
\subsection{Procedure}
\noindent Each experiment is prepared by filling the basin with $\mathrm{TiO}_{2}$-water dispersion until it wets the entire substrate. After removing some liquid, the system is in a metastable state that enables the existence of stable holes. In order to create a hole, the film is ruptured at the center using an impinging jet of gas. By adding liquid again, the hole size can be adjusted for the subsequent experiments that are conducted in the following way:
\begin{enumerate}
    \item At a specific frequency ($\qty{100}{\hertz}\leq f\leq\qty{200}{\hertz}$) the film is vibrated with the maximum velocity amplitude, that is $\qty{65}{\milli\metre\per\second}$ for all frequencies. 
    \item Over a duration of at least 15 seconds, images are recorded from the top at 1 frame per second (fps), while the height profile data are recorded with the laser sensor.
    \item The velocity amplitude is reduced by $\qty{6.5}{\milli\metre\per\second}$.
    \item Step 2 and 3 are repeated until the velocity amplitude reaches zero, which serves as the static reference point for the previous experiments.
\end{enumerate}
Repeating this protocol at different initial hole sizes and at different frequencies gives a set of experimental data with three parameter variations: amplitude, frequency and hole size. One set of experiments where the amplitude is varied over the entire range is depicted in Fig.~\ref{fig:overview_wave_energy}.
\subsection{Evaluation}
\subsubsection{Hole Radius}
\noindent From the image data we extract the diameter of the hole. We locate the hole of the first image of a sequence manually. Based on the \textit{adaptiveThreshold}-method of the python library \textit{openCV}, we then automatically determine the area of the exposed substrate in all subsequent images 
\footnote{\url{https://opencv.org/}}. 
Assuming the hole to be circular the exposed area is used to determine the radius of the hole. The automatic detection of the exposed area is exemplified on the left side of Fig.~\ref{fig:visual_setup}.\\
\subsubsection{Wave Energy Measurement}
\begin{figure}
	\includegraphics[width=0.92\textwidth]{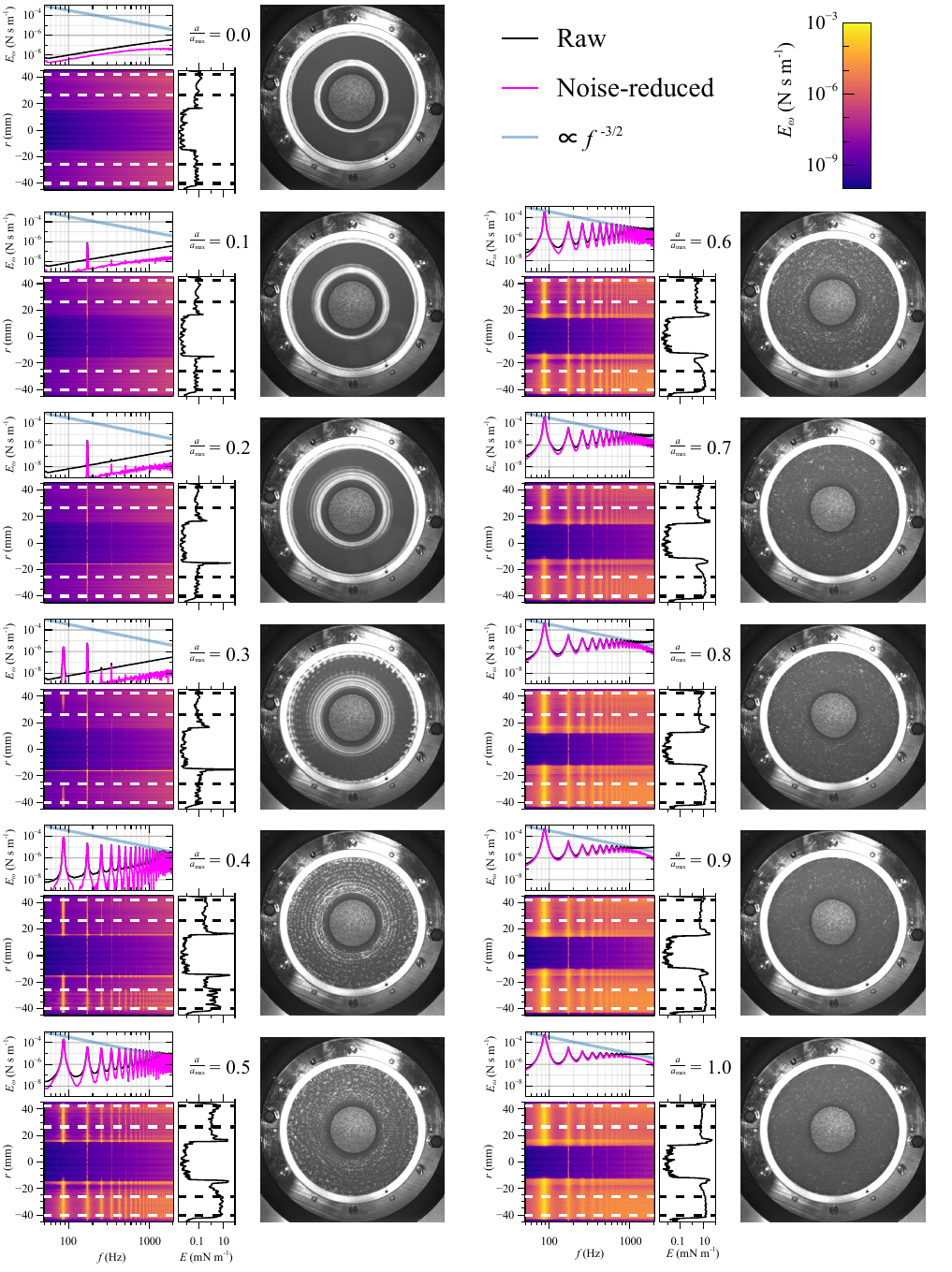}
	\caption{\justifying Overview of wave energy measurements for varying exciter amplitude. The maximum velocity amplitude is $\qty{65}{\milli\metre\per\second}$ (bottom right). The exciter frequency is $\qty{170}{\hertz}$. The color plot shows the energy spectrum at different positions along the vertical diameter of the basin corresponding to the image next to it. The spatial average of the energy spectrum is shown above. The frequency integrated energy spectrum (i.e. wave energy) as a function of the position is shown on the right.}
	\label{fig:overview_wave_energy}
\end{figure}

\noindent On average, the energy of capillary waves with sufficiently small amplitude ($|\nabla\eta|\ll 1$) contains equal contributions from kinetic and potential energy \cite{Galtier.2021, Deike.2014}. The average potential energy per unit area of a unidirectional standing water wave is 
\begin{equation}
E_{\mathrm{pot}} = E/2 = \frac{1}{2}\rho g(1+Bo^{-1})\cdot\left\langle\eta^2\right\rangle_t.
\end{equation}
For the case of a spectrum of interfering waves, we can integrate the energy spectrum of the waves to obtain the potential wave energy
\begin{equation}
E_{\mathrm{pot}} = E/2 = \frac{1}{2}\rho g\int_{0}^{\infty}(1+Bo^{-1})\cdot\left\langle\eta_{\omega}^2\right\rangle_t\mathrm{d}\omega = \frac{1}{2}\int_{0}^{\infty} E_{\omega}\mathrm{d}\omega.
\end{equation}
For capillary waves, weak wave turbulence theory predicts a scaling of $E_{\omega}\propto f^{-3/2}$ that makes the expression for $E$ converge, even when not considering a viscous lengthscale that cuts off the high frequency tail of the spectrum~\cite{Zakharov.1967}. In the experiments we cannot resolve the energy spectrum up to this point. This is due to the limited sensitivity of the sensor: the laser sheet thickness is $\qty{610}{\micro\metre}$ and the equivalent pixel width is about $\qty{300}{\micro\metre}$. This means that the sensitivity of the sensor decreases rapidly around wavelengths of $\qty{610}{\micro\metre}$ or frequencies of $\qty{1414}{\hertz}$, using the inviscid dispersion relation. Additionally, the measurement of $\left\langle\eta_{\omega}^2\right\rangle_t$ is subject to white noise. Therefore, the noise in $E_{\omega}$ scales as ${Bo^{-1}\propto f^{4/3}}$. These two effects (decreasing sensitivity and increasing noise) make for a decreasing signal-to-noise ratio with increasing frequency. In order to not overestimate the wave energy, we subtract the noise based on an estimation at the high-frequency tail where the signal mainly consists of noise. The cut-off frequency is then set to $\qty{1414}{\hertz}$, above which we cannot expect to properly measure capillary waves anymore. We can make a conservative estimate on how much wave energy is cut off by this procedure. We expect the error to be largest for large exciter amplitudes, since the high-end frequency spectrum is more developed in this case. Consider the measurement described on the bottom right in Fig. \ref{fig:overview_wave_energy}. We have $E_{\omega}(f=\qty{1414}{\hertz})\approx \qty{3e-6}{\newton\second\per\metre}$. The measured wave energy is $E_{\mathrm{measure}}\approx \qty{10}{\newton\per\metre}$. Assuming that $E_{\omega}\propto f^{-3/2}$, we have
\begin{equation}
\begin{aligned}
\frac{E_{\mathrm{cutoff}}}{E_{\mathrm{measured}}} < \frac{1}{E_{\mathrm{measured}}} \int_{\qty{1414}{\hertz}}^{\infty} E_{f}\quad\mathrm{d}f = \frac{1}{E_{\mathrm{measured}}} \int_{\qty{1414}{\hertz}}^{\infty}\frac{\qty{3e-6}{\newton\second\per\metre}}{(\qty{1414}{\hertz})^{-3/2}}f^{-3/2}\mathrm{d}f = \\\\\frac{\qty{3e-6}{\newton\second\per\metre}\cdot 2\cdot \qty{1414}{\hertz}}{E_{\mathrm{measured}}}<0.001.
\end{aligned}
\end{equation}
For this reason, we estimate that the error from disregarding the energy spectrum above $\qty{1414}{\hertz}$ is small for all experiments.\\
\hspace{6pt} The power spectrum of the liquid film itself $\left\langle\eta_{\omega}^2\right\rangle_t$ is obtained by means of the spectral density estimation method \textit{signal.welch} as part of the signal processing toolbox of the Python library \textit{SciPy}
\begin{savenotes}
\footnote{\url{https://scipy.org}}
\end{savenotes}
.
\clearpage
\section{Bayesian Estimation of the Effective Capillary Length}

\subsection{Bayesian Model $p(\Theta|D)$}
The Bayesian model consists of two likelihood functions describing the static and vibrated cases as well as prior distributions:
\begin{align}
    p(\Theta|D) &\propto \mathcal{L}(\Theta|D)p(\Theta)\\
    \log{\mathcal{L}(\Theta|D)} &= \log{\mathcal{L}^{\mathrm{static}}(\Theta^{\mathrm{static}}|D)} + \log{\mathcal{L}^{\mathrm{vibrated}}(\Theta^{\mathrm{vibrated}}|D)} = \\
            &\sum_{i,j}\left (
                -N_{i,j}/2 \log{2\pi\sigma_{i,j}^2} - 1/(2\sigma_{i,j}^2)\sum((D_{i,j}-\hat D(\Theta^{\mathrm{static}}))^2\right ) + \notag\\
            &\sum_{i,j,k}\left (
                -N_{i,j,k}/2 \log{2\pi\sigma_{i,j,k}^2} - 1/(2\sigma_{i,j,k}^2)\sum((D_{i,j,k}-\hat D(\Theta^{\mathrm{vibrated}}))^2\right )\notag\\
    \Theta^{\mathrm{static}} &= (\theta_{i}, l^\mathrm{c}_i, \Delta V_{i,j})\\
    \Theta^{\mathrm{vibrated}} &= (\theta_{i}, l^\mathrm{c}_i, \alpha_{i,j,k}, \Delta V_{i,j})
\end{align}
In the static cases, the data $D_{i,j}$ is indexed per substrate or day $i$ (one substrate per day) and volume $j$.
The vibrated cases are additionally indexed per amplitude of the shaker $k$.
The measurement noise is modeled with a normal distribution with zero mean and a standard deviation of $\sigma_{i,j}$ or $\sigma_{i,j,k}$.
For the simulated data $\hat D$ see below.
The uncertain parameters are the substrate contact angles $\theta_i$, the static capillary lengths $l_i^\mathrm{c}$, volume errors $\Delta_{i,j}$ and, most importantly, parameters $\alpha_{i,j,k}=l_{i,j,k}/l_{i}^\mathrm{c}$, where $l_{i,j,k}$ are the effective capillary lengths.
The parameters of the Bayesian model, most notably the effective capillary length, are constrained by the observable diameter of the holes in the liquid films $D$ and $\hat D$.

\subsection{Simulated Observations $O(\mathit{YL}(\Theta))$}
In the experiment, the hole diameters are extracted from top view images, see Fig. \ref{fig:visual_setup}.
Due to the possible curvature of the gas-liquid interface, the measured hole diameter does not necessarily correspond to the hole diameter right at the substrate, but is merely the projection of the liquid from top view.
The observation operator $O$ mimics this restriction from the experiments:
\begin{align}
    \hat D(\Theta) &= O(\mathit{YL}(\Theta)) \text{ (simulated observations)}\\
    O(r, z, \phi) &= \min(r) \text{ (observation operator)}\\
    \mathit{YL}(\theta, l, V) &= (r, z, \phi) \text{ (forward model)}\;,
\end{align}
with $l = l^\mathrm{c} \alpha$, $\alpha = 1$ for the static case, and $V = \hat V + \Delta V$.

\subsection{Forward Model $\mathit{YL}(\theta, l, V)$}
The forward model $\mathit{YL}(\theta, l, V)$ (Young-Laplace system),
\begin{equation}
\begin{array}{rrclcl}
        \displaystyle\min_{\Delta p_0} & & \multicolumn{2}{l}{|\Delta p_0|^2_2} \\
        \textrm{s.t.}& \frac{d \phi}{d s} &=& -\frac{\sin\phi}{r}+z-\Delta p_0 \\
        & \frac{d r}{d s} &=& \cos\phi, \frac{d z}{d s} = \sin\phi \\
        & z(s_\mathrm{S}) &=& 0, \phi(s_\mathrm{S}) = \theta, r(s_\mathrm{W}) = r_{s_\mathrm{W}} , z(s_\mathrm{W}) = z_{s_\mathrm{W}} \\
        & \int 2 \pi r z \diff{r}{s} \mathrm{d}s &=& V \\
        & r &\geq& 0, z \geq 0\;,
\end{array}
\end{equation}
is implemented as an ODE-constraint optimization problem in \textit{Pyomo}
\footnote{\url{https://www.pyomo.org/}}
and solved via \textit{Ipopt}
\footnote{\url{https://github.com/coin-or/Ipopt}}
.

\subsection{Priors $p(\Theta)$}
We use relatively weak priors, i.e., large standard deviations, centered around the expected values for the nuisance parameters $\theta$, $l^\mathrm{c}$, and $\Delta V$. 
The expected values are based on separate, independent measurements of contact angle and capillary length. 
For the most interesting parameter $\alpha$, we use a uniform distribution between some reasonable bounds to not inform the inference too much.
\begin{align}
    \theta_{i} &\sim \mathcal{N}(150, 10)\si{\degree}\\
    l^\mathrm{c}_i &\sim \mathcal{N}(2.73, 0.3)\si{\milli\meter}\\
    \Delta V_{i,j} &\sim \mathcal{N}(0, 4)\si{\milli\liter}\\
    \alpha_{i,j,k} &\sim \mathcal{U}(0.915, 1.025) 
\end{align}

\subsection{Surrogate Model $\hat{\mathit{YL}}(\theta, l, V)$}
\begin{figure}[!h]
    \centering
    \includegraphics{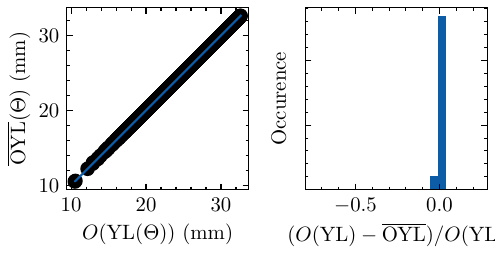}
    \caption{\justifying Comparison of the full model $O(\mathit{YL}(\Theta))$ with the Gaussian process regression surrogate $\mathit{\overline{OYL}}(\Theta)$ (1500 training and test points each). The error of the surrogate is very small for the whole range of parameters.}
    \label{fig:comp_surr}
\end{figure}

To reduce the computing time to approximate $p(\Theta|D)$ we replace $\mathit{YL}$ with a highly accurate and fast surrogate model $\mathit{\hat YL}(\theta, l, V)$.
The training and test data is generated via simulation using $\mathit{YL}(\theta, l, V)$.
In total, 3000 tuples $(\theta, l, V)$ are simulated in a parameter space spanned by:
\begin{align}
    \theta &\text{ in } [130,170]\unit{\degree}\;,\\
    V &\text{ in } [10, 30]\unit{\milli\liter}\;,\\
    l &\text{ in } [2, 4]\unit{\milli\meter}\;.
\end{align}
The 3000 points are allocated via active learning using the software \textit{adaptive}
\footnote{\url{https://adaptive.readthedocs.io}}
.
Afterwards, Gaussian process regression provided by~\textit{SMT}
\footnote{\url{https://smt.readthedocs.io}}
is performed with 1500 randomly chosen training points.
This results in an excellent predictive accuracy above 0.99 using the remaining 1500 test training points.
For a visual comparison of $\mathit{\hat{YL}}$ with $\mathit{YL}$, see~fig.\ref{fig:comp_surr}.

\subsection{Solution Procedure for $l(i,j,k)=\alpha(i,j,k)l^\mathrm{c}_i$}
\begin{figure}
    \centering
    \includegraphics{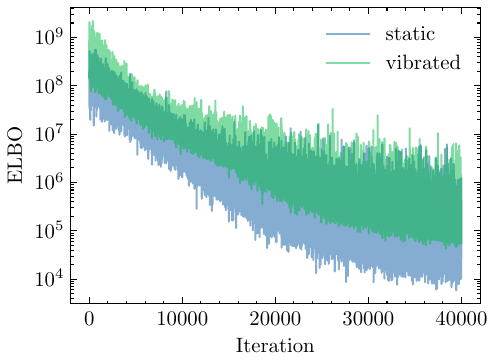}
    \caption{\justifying Evidence lower bound (ELBO) from ADVI per iteration for the static and vibrated cases. In both cases, the ELBO stabilizes at a very small value indicating convergence.}
    \label{fig:advi_conv}
\end{figure}

\begin{figure}
    \centering
    \includegraphics[scale=0.9]{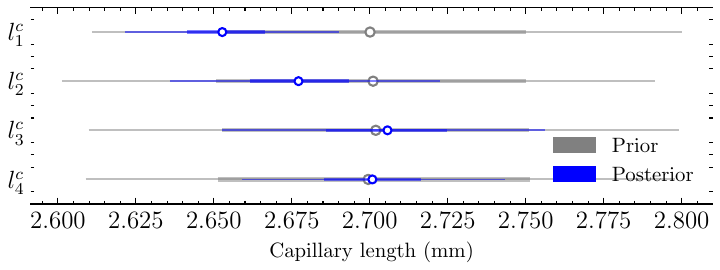}
    \includegraphics[scale=0.9]{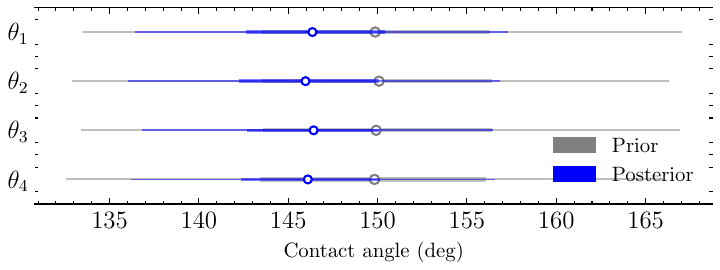}
    \includegraphics[scale=0.9]{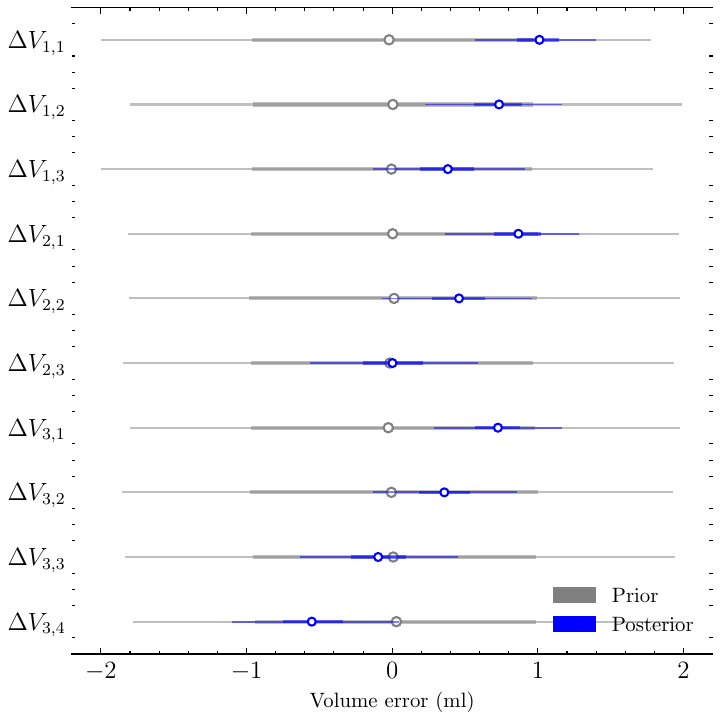}
    \caption{
    \justifying Forrest plots of prior and posterior distributions of $l^c_i$ (top), $\theta_i$, and $\Delta V_{i_j}$ (bottom). 
    The wide priors (gray) get tightly constricted around mean values between 2.65 and 2.7 \si{\milli\metre} (top, indicating fluctuations of, likely, temperature and humidity), around \SI{146}{\degree} (indicating good reproducibility of the substrate modification), and between -1 and +1 \si{\milli\liter} (bottom, correlations between volume errors are likely due to evaporation between experiments).}
    \label{fig:post}
\end{figure}

The Bayesian model is implemented in the Python package \textit{PyMC}
\footnote{\url{https://www.pymc.io}}
.
In a first step, only experimental data from static experiments are used to approximate $p(\Theta_\mathrm{static}|D)$.
In a second step, the posterior $p(\Theta_\mathrm{static}|D)$ is used as the prior for the approximation of $p(\theta_\mathrm{vibrated}|D)$ together with the experimental data from the vibrated experiments.
Each approximation is performed using full rank automatic differentiation variational inference (ADVI) with 30000 iterations to ensure convergence, see figure~\ref{fig:advi_conv}.
For the resulting posterior of $l(i,j,k)$ and the nuisance parameters, see the main text and figure~\ref{fig:post}, respectively.

\clearpage
\section{Theory}
\subsection{Radiation Pressure of Pseudo-Chaotic Capillary Waves}
\begin{figure}[h]
	\includegraphics[width=0.3\textwidth]{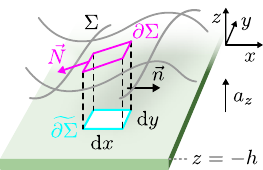}
	\caption{\justifying Control volume in the deformed, infinitely extended film considered to compute $S_{\mathrm{rad}}$. Only forces that act in horizontal direction are shown.}
	\label{fig:problem_geoemtry}
\end{figure}

\noindent Consider an infinitely extended liquid film on a horizontal substrate. A section of this film is shown in Fig. \ref{fig:problem_geoemtry}.
The substrate is vibrated in $z$-direction. Fixing our frame-of-reference to the substrate, this manifests itself in an oscillating acceleration $\vec{a} = \vec{e_z}(-g-a\cos(\omega t))$ where $\omega$ is the (angular) excitation frequency, $g$ is the gravitational acceleration, and $a$ is the acceleration amplitude of the exciter. We define a time average and a spatial average
\begin{align}
    \left\langle\cdot\right\rangle_{t} &= \frac{1}{T}\int^{T}_{0}\cdot\quad\dif t\\
    \mathrm{and}\qquad\left\langle\cdot\right\rangle_{x} &= \frac{1}{X}\int^{X}_{0}\cdot \quad\dif x
\end{align}
where $T$ and $X$ are large compared to the period and wavelength of the first subharmonic wave. Spatial averaging in the $y$-direction is analogous to that in the $x$-direction. Consider an arbitrary field quantity $\phi(x,y,z,t)$. We assume homogeneous and isotropic chaotic waves, so we have
\begin{align}
\left\langle\left\langle\left\langle \phi(x,y,z,t) \right\rangle_{t}\right\rangle_{x}\right\rangle_{y} = \left\langle \phi(x,y,z,t) \right\rangle_{t}.
\end{align}
We parameterize the liquid surface $\Sigma$ by a height function $z = \eta(x,y,t)$ that describes the displacement relative to the static film for which we have $z=0$. For the time-averaged height function we have
\begin{align}
\left\langle\left\langle\left\langle \eta(x,y,t) \right\rangle_{t}\right\rangle_{x}\right\rangle_{y} = \left\langle \eta(x,y,t) \right\rangle_{t} = 0.\label{eq:eta}
\end{align}
Neglecting viscous stresses, in Cartesian coordinates the conservation of momentum for a fixed volume containing a column of liquid is described by
\begin{align}
\frac{\partial}{\partial t}\int\displaylimits_{V}\rho u_{i}\dif v = -\int\displaylimits_{\partial V}\left(\rho u_i u_j + p\delta_{ij}\right)n_j \dif A +\int\displaylimits_{\partial\Sigma}\sigma N_i \dif s + \int\displaylimits_{V}\rho a_i \dif v.
\end{align}
The control volume $V$ is bounded by $\partial V$. The outward unit normal on $\partial V$ is $\vec{n}$. The intersect of the gas liquid interface $\Sigma$ and $\partial V$ is denoted by $\partial\Sigma$. The unit normal on $\partial \Sigma$ that is tangent to $\Sigma$ and outward directed is $\vec{N}$. The surface tension $\sigma$ acts along $\partial\Sigma$ and parallel to $\vec{N}$. The velocity components  are denoted by $u_i$, the mass density by $\rho$ and pressure by $p$. Now we examine the momentum transport in an arbitrary horizontal direction. We choose $i=x$ ($(u_x,u_y,u_z)=(u,v,w)$) without loosing generality and get
\begin{align}
    \frac{\partial}{\partial t}\int\displaylimits_{V}\rho u\dif v = -\int\displaylimits_{\partial V}\left(\rho uv n_y+\rho u^2n_x+pn_x\right)\dif A + \int\displaylimits_{\partial\Sigma}\sigma N_x\dif s.\label{eq:int_momentum_balance_2}
\end{align}
Assuming that the control volume is infinitisimally small in $x$- and $y$-direction and using the small-angle approximation (i.e. $\partial\eta/\partial x\ll 1$ and $\partial\eta/\partial y\ll 1$) we have
\begin{align}
    N_x &= \cos{\left(\arctan{\left(\frac{\partial\eta}{\partial x}\right)}\right)} = \cos{\left(\frac{\partial\eta}{\partial x}\right)}\label{eq:surf_kin_1}\\
    \mathrm{and}\qquad \dif s &= \frac{1}{\cos{\left(\arctan{\left(\frac{\partial\eta}{\partial y}\right)}\right)}}\dif y = \frac{1}{\cos{\left(\frac{\partial\eta}{\partial y}\right)}}\dif y\label{eq:surf_kin_2}
\end{align}
for the integrations where $\partial\Sigma$ is orthogonal to the $y$-axis. For the case where $\partial\Sigma$ is orthogonal to the $x$-axis, we have $N_{x}=0$.
For a static system, both sides of the equation \ref{eq:int_momentum_balance_2} are zero. After taking the time average, both sides also vanish for the dynamic system, and using equations \ref{eq:surf_kin_1} and \ref{eq:surf_kin_2} we get
\begin{align}
    0 = \left\langle\int\displaylimits_{\partial V}\left(\rho uv n_y+\rho u^2n_x+pn_x\right)\dif A\right\rangle_t - \left\langle\int\displaylimits_{\widetilde{\partial\Sigma}}\sigma \frac{\cos{\left(\frac{\partial\eta}{\partial x}\right)}}{\cos{\left(\frac{\partial\eta}{\partial y}\right)}}n_x\dif \tilde{s}\right\rangle_t.
\end{align}
We now reformulate the integration by $\dif A = \dif z\dif \tilde{s}$. Then we get
\begin{align}
    0 = \int\displaylimits_{\widetilde{\partial\Sigma}}\left\langle\int_{-h}^{\eta}\left(\rho uv n_y+\rho u^2n_x+pn_x\right)\dif z\right\rangle_t\dif \tilde{s} - \int\displaylimits_{\widetilde{\partial\Sigma}}\left\langle\sigma \frac{\cos{\left(\frac{\partial\eta}{\partial x}\right)}}{\cos{\left(\frac{\partial\eta}{\partial y}\right)}}n_x\right\rangle_t\dif \tilde{s},
\end{align}
where we have used the fact that the time average and the horizontal spatial integration commute. For isotropic waves, the first shear-stress term in the first integral is zero after integration along $z$ and time averaging. The effect of the remaining terms can be subsumed in a pressure term
\begin{align}
    0 = \int\displaylimits_{\widetilde{\partial\Sigma}}  \left\langle\Phi_{\mathrm{dyn}}\right\rangle_t n_x \dif\tilde{s}.
\end{align}
Thus, we define the time-averaged, height-integrated momentum flux $\left\langle\Phi_{\mathrm{dyn}}\right\rangle_t$ by
\begin{align}
   \left\langle\Phi_{\mathrm{dyn}}\right\rangle_t = \left\langle\:\int_{-h}^{\eta}\left(\rho u^2+ p\right) \dif z -\sigma\frac{\cos{\left(\frac{\partial\eta}{\partial x}\right)}}{\cos{\left(\frac{\partial\eta}{\partial y}\right)}}\right\rangle_t.
\end{align}
We define the radiation pressure as the access momentum flux compared to the static system by
\begin{equation}
\begin{aligned}
    S_{\mathrm{rad}} \coloneqq& \left\langle\Phi_{\mathrm{dyn}}\right\rangle_t
    -\left(\int_{-h}^{0}-\rho gz\dif z-\sigma\right) \\=& \left\langle\:\int_{-h}^{\eta}\left(\rho u^2+ p\right) \dif z -\sigma\frac{\cos{\left(\frac{\partial\eta}{\partial x}\right)}}{\cos{\left(\frac{\partial\eta}{\partial y}\right)}}\right\rangle_t -\left(\int_{-h}^{0}-\rho gz\dif z-\sigma\right).\label{eq:def_rad_pressure}
\end{aligned}
\end{equation}
\noindent The main problem with the right-hand-side of this equation is the fact that we cannot draw the time-average under the first integral, which has a time-dependent upper bound $\eta$.
When splitting up the integration domain for the pressure $p$ in equation \ref{eq:def_rad_pressure} and rearranging it we obtain
\begin{equation}
\begin{aligned}
    S_{\mathrm{rad}} = &\left\langle\int_{-h}^{\eta}\rho u^2\dif z\right\rangle_t+\int_{-h}^{\eta_{\mathrm{min}}}\left(\left\langle p\right\rangle_t+\rho gz\right)\dif z + \left\langle\int_{\eta_{\mathrm{min}}}^{\eta} p \dif z\right\rangle_t \\
    &+\int_{\eta_{\mathrm{min}}}^{0}\rho gz\dif z + \sigma\left\langle1-\frac{\cos{\left(\frac{\partial\eta}{\partial x}\right)}}{\cos{\left(\frac{\partial\eta}{\partial y}\right)}}\right\rangle_t,\label{eq:rad_stress_little_simpler}
\end{aligned}
\end{equation}
where $\eta_{\mathrm{min}} = \min{(\eta)}$. Then, we are able to draw the time average into the second integral. The last surface tension-dependent term vanishes for the case of isotropic chaotic waves. Note that this is not the only place where surface tension is at play: it is still present in the pressure integral in the form of the Laplace pressure.\\
To add information on the pressure $p$, we use a time averaged vertical momentum balance over a fixed control volume, similar as the one used so far. We consider a control volume with its bottom surface located at an arbitrary vertical position $z$ (c.f. \cite{LonguetHiggins.1964}). We get
\begin{equation}
\begin{aligned}
    0 = &\left\langle p +\rho w^2 \right\rangle_t +\left\langle \rho (g+a\cos(\omega t))(z-\eta)\right\rangle_t\\
    &+\left\langle\frac{\partial}{\partial x}\int_{z}^{\eta}\rho uw\dif\tilde{z}+\frac{\partial}{\partial y}\int_{z}^{\eta}\rho vw\dif\tilde{z}\right\rangle_t+\sigma\left\langle\kappa\right\rangle_t\label{eq:liquid_column_pressure},
\end{aligned}
\end{equation}
where the first term is the momentum flux into the control volume through its bottom, the second one is the momentum production inside the volume (area specific weight of the liquid column), the third represents fluxes through the sides and the last one is the momentum flux through the top. For the latter, the contribution of the surface tension along $\partial\Sigma$ was rewritten as a surface integral over the curvature $\kappa$, localized now at a specific $x$-$y$-point. $\kappa$ is defined in such a way that $\kappa<0$ when the liquid film is locally convex. Based on the assumption of homogeneous and isotropic chaotic waves, we have
\begin{align}
    \left\langle\frac{\partial}{\partial x}\int_{z}^{\eta}\rho uw\dif\tilde{z}+\frac{\partial}{\partial y}\int_{z}^{\eta}\rho vw\dif\tilde{z}\right\rangle_t = 0.
\end{align}
Motivated by the fact that the frequency of most of the waves differs from the excitation frequency $\omega $ (mostly subharmonic waves), we assume
\begin{align}
    \left\langle\cos(\omega t)\eta\right\rangle_t = 0.
\end{align}
With that, equation \ref{eq:liquid_column_pressure} simplifies to
\begin{align}
    0 = \left\langle p\right\rangle_t + \left\langle \rho w^2\right\rangle_t + \sigma\left\langle\kappa\right\rangle_t + \rho gz.\label{eq:pressure}
\end{align}

\noindent Using equation \ref{eq:pressure}, we can simplify the second integral in equation \ref{eq:rad_stress_little_simpler}:
\begin{align}
    \int_{-h}^{\eta_{\mathrm{min}}}\left\langle p\right\rangle_t+\rho gz\dif z = -\int_{-h}^{\eta_{\mathrm{min}}}\left\langle\rho w^2\right\rangle +\sigma\left\langle\kappa\right\rangle_t\dif z = -\int_{-h}^{\eta_{\mathrm{min}}}\left\langle\rho w^2\right\rangle_t\dif z.
\end{align}
Inserting this in equation \ref{eq:rad_stress_little_simpler} we get
\begin{align}
    S_{\mathrm{rad}} = \left\langle\int_{-h}^{\eta}\rho u^2\dif z\right\rangle_t-\int_{-h}^{\eta_{\mathrm{min}}}\rho \left\langle w^2\right\rangle_t\dif z+\left\langle\int_{\eta_{\mathrm{min}}}^{\eta}p\dif z\right\rangle+ \int_{\eta_{\mathrm{min}}}^{0}\rho gz\dif z.\label{eq:rad_stress_3}
\end{align}
Close to the interface velocity gradients in z-direction are small, so the last two integrals can be simplified to
\begin{equation}
\begin{aligned}
    \left\langle\int_{\eta_{\mathrm{min}}}^{\eta}p\dif z\right\rangle_t + \left\langle\int_{\eta_{\mathrm{min}}}^{0} \rho gz\dif z\right\rangle_t &= \left\langle\int_{\eta_{\mathrm{min}}}^{\eta}\rho g(\eta -z) -\sigma\kappa\dif z\right\rangle_t -\frac{1}{2}\rho g\eta_{\mathrm{min}}^2\\ &= \left\langle\frac{1}{2}\rho g\eta^2\right\rangle_t-\left\langle\sigma\kappa\eta\right\rangle_t.
\end{aligned}
\end{equation}
Inserting this in equation \ref{eq:rad_stress_3} we get
\begin{align}
    S_{\mathrm{rad}} = \left\langle\int_{-h}^{\eta}\rho u^2\dif z\right\rangle_t - \int_{-h}^{\eta_{\mathrm{min}}}\rho\left\langle w^2\right\rangle_t\dif z + \frac{1}{2}\rho g\left\langle\eta^2\right\rangle_t - \sigma\left\langle\kappa\eta\right\rangle_t.\label{eq:rad_stress_simpler}
\end{align}
To proceed, we recall that the (area specific, total) wave energy is given by
\begin{align}
    E = \rho g\left\langle\eta^2\right\rangle_t\left(1+Bo^{-1}\right),
\end{align}
with the Bond number
\begin{align}
    Bo = \frac{\rho g}{\sigma k^2}.
\end{align}
It describes the ratio of capillary and gravity forces, where the length scale enters via the wave number $k$. For a standing wave of amplitude $|\eta_{\mathrm{min}}|$ the rightmost term in equation \ref{eq:rad_stress_simpler} is
\begin{align}
    - \sigma\left\langle\kappa\eta\right\rangle_t = \frac{\sigma k^2}{2}\left\langle\cos^2(\omega t)a^2\right\rangle_t = \frac{\sigma k^2}{4}\eta_{\mathrm{min}}^2 = \frac{\sigma k^2}{2}\left\langle\eta^2\right\rangle_t.
\end{align}
With that, we obtain for the radiation pressure
\begin{align}
    S_{\mathrm{rad}} &=  \left\langle\int_{-h}^{\eta}\rho u^2\dif z\right\rangle_t - \int_{-h}^{\eta_{\mathrm{min}}}\rho\left\langle w^2\right\rangle_t\dif z + \frac{1}{2}\rho g\left\langle\eta^2\right\rangle_t + \frac{\sigma k^2}{2}\left\langle\eta^2\right\rangle_t \\
    &=  \left\langle\int_{-h}^{\eta}\rho u^2\dif z\right\rangle_t - \int_{-h}^{\eta_{\mathrm{min}}}\rho\left\langle w^2\right\rangle_t\dif z + \frac{E}{2}.
\end{align}
Assuming equipartition of the time averaged kinetic and potential kinetic energy in the liquid film, i.e.
\begin{align}
    \frac{E}{2} = \left\langle E_{\mathrm{pot}}\right\rangle_t = \left\langle E_{\mathrm{kin}}\right\rangle_t = \frac{\rho}{2}\left\langle\int_{-h}^{\eta} u^2+v^2+w^2\dif z\right\rangle_t,
\end{align}
we find that the radiation pressure is dominated by the surface term and not by the inertial term:
\begin{align}
    \frac{\frac{E}{2}}{ \left\langle\int_{-h}^{\eta}\rho u^2\dif z\right\rangle_t - \int_{-h}^{0}\rho\left\langle w^2\right\rangle_t\dif z} = \frac{\left\langle\int_{-h}^{\eta} u^2+v^2+w^2\dif z\right\rangle_t}{2 \left\langle\int_{-h}^{\eta} u^2\dif z\right\rangle_t - \int_{-h}^{0}\left\langle w^2\right\rangle_t\dif z} \gg 1.\label{eq:est_inertia_surface}
\end{align}
Neglecting the inertial term, we obtain for the radiation pressure
\begin{align}
    S_{\mathrm{rad}} = \frac{E}{2} = \frac{1}{2}\rho g\left\langle\eta^2\right\rangle_t\left(1+Bo^{-1}\right).
\end{align}


\bibliography{literature}